\documentclass[10pt,aps,prd,nofootinbib,superscriptaddress,twocolumn,preprintnumbers,balancelastpage]{revtex4-2}

\usepackage{xcolor}
\usepackage{float}
\usepackage{amssymb,amsmath} 

\usepackage[colorlinks=true,urlcolor=blue,anchorcolor=blue,citecolor=blue,filecolor=blue,linkcolor=blue,menucolor=blue,linktocpage=true,pdfproducer=medialab,pdfa=true]{hyperref}
\usepackage[capitalize]{cleveref}

\usepackage{booktabs}

\usepackage{bm}
\usepackage{setspace}

\def\equationautorefname~#1\null{Eq.\,(#1)\null}

\makeatletter\g@addto@macro\bfseries{\boldmath}\makeatother

\newcommand{\lag}{\mathcal{L}}
\newcommand{\abs}[1]{\left\lvert #1 \right\rvert}

\begin{document}

\preprint{CERN-TH-2025-157}

\title{Flavor hierarchies with nonminimal irreducible representations}

\author{Hannah Banks}
\affiliation{Department of Applied Mathematics and Theoretical Physics,
University of Cambridge, Wilberforce Road, Cambridge, CB3 0WA, UK}
\affiliation{Center for Cosmology and Particle Physics, Department of Physics, New York University, New York, NY 10003}

\author{Graeme Crawford}
\affiliation{School of Physics and Astronomy, University of Glasgow, Glasgow G12 8QQ, UK}

\author{Matthew McCullough}
\affiliation{Theoretical Physics Department, CERN, 1211 Geneva 23, Switzerland}

\author{Dave Sutherland}
\affiliation{School of Physics and Astronomy, University of Glasgow, Glasgow G12 8QQ, UK}


\begin{abstract}
\begin{center}
{\bf Abstract}
\vspace{-7pt}
\end{center}
We propose a new class of flavour models in which the spurion which breaks Standard Model flavour symmetries transforms in a non-minimal representation. Hierarchies in fermion masses, which arise from multiple insertions of this spurion, may be generated in a technically natural, accidental manner, from a handful of untuned $\mathcal{O}(1)$ elements in the UV. This relies explicitly on the non-Abelian nature of the symmetry, distinguishing it from standard Froggatt-Nielsen-like scenarios.  The pattern of flavour violating operators at dimension-6 can radically differ from previously considered scenarios, and emphasises the need for a broad flavour programme across all generations.
\end{abstract}

\maketitle

\section{Introduction}
\label{sec:introduction}

During the 19$^{\text{th}}$ century it was recognized that, arranged by mass and chemical properties, the chemical elements exhibit intriguing patterns.  The modern manifestation of these patterns, Mendeleev's periodic table, appeared in 1869, long before the advent of quantum mechanics, the discovery of the atomic nucleus, and so much more that is necessary to explain those patterns.  It took many decades to answer the `Why?' of the periodic table.


The analogous Standard Model (SM) flavour puzzle concerns the structure of the SM Yukawa couplings, which are the only interactions that explicitly break a number of global flavour symmetries.  Understanding their origin thus ultimately boils down to understanding \textit{why} these symmetries are broken, across families, in a hierarchical manner.  Harnessing the power of spurion analysis, the Yukawas can be interpreted as spurions transforming in the $\mathbf{3} \times \overline{\mathbf{3}}$ of a pair of SU$(3)$ global symmetries.  If we can understand the origin of these hierarchical spurions, we may begin to resolve the flavour puzzle.

Within this perspective the Yukawas have been subject to ceaseless scrutiny.  The Minimal Flavour Violation (MFV) \cite{Chivukula:1987py,Hall:1990ac,DAmbrosio:2002vsn,Buras:2000dm,Cirigliano:2005ck,Isidori:2012ts} ansatz aims not to explain their origin but instead the implications for experimental signatures of the microscopic world beyond the SM (BSM) under the assumption that the Yukawas correspond to the only sources of flavour violation in the deep UV.  A similar ansatz assumes an approximate U$(2)$ symmetry in the flavour sector, such that the third generation are separated in a form that goes beyond MFV \cite{Pomarol:1995xc,Barbieri:1995uv,Barbieri:1996ww,Barbieri:1997tu,Feldmann:2008ja,Kagan:2009bn,Barbieri:2012uh,Barbieri:2011ci,Fuentes-Martin:2019mun,Greljo:2023bix,Greljo:2024zrj,Lizana:2024jby}.  Further generalisations have also been studied \cite{Agashe:2005hk,Egana-Ugrinovic:2018znw}.  The Froggatt-Nielsen (FN) approach \cite{Froggatt:1978nt,Leurer:1992wg,Leurer:1993gy,Fedele:2020fvh,Smolkovic:2019jow} aims to explain the hierarchical patterns as following from a small spurion charged under a U$(1)_\text{FN}$ symmetry, such that the larger the charge of a fermion, the more spurion insertions are required to generate the Yukawa, hence the greater the mass suppression.  Finally, it is worth noting that new physics sectors may be flavour universal, such that flavour-violation is greatly suppressed, arising only through SM RG contributions. For a recent review of the landscape of flavour model building efforts see Refs.~\cite{Altmannshofer:2024hmr,Altmannshofer:2025rxc}.

In this work our approach is to take the non-Abelian nature of the flavour symmetries seriously and explore the UV landscape more systematically.  We assume that the flavour symmetries are broken in the UV, whose IR effects may be captured by spurions transforming in irreducible representations (irreps) of the flavour symmetries, with non-zero values.  Where our work departs significantly from preceding literature is that we consider spurions which are in `non-minimal' irreps of the flavour symmetries.

As we shall demonstrate, this approach gives rise to behaviour not captured in traditional frameworks, generating novel implications for phenomenology and model building.  The goal is not to motivate a particular scale for flavour, but rather to map, from a symmetry perspective, the universality class of possible microscopic origins of the observed Yukawa hierarchies and their phenomenological consequences.  

Calling the spurion $\mathcal{Y}$, to generate a Yukawa, $y$, one must be able to construct a $\mathbf{3} \times \overline{\mathbf{3}}$ under $\mathrm{SU}(3)\times\mathrm{SU}(3)$ out of $\mathcal{Y}$ and any other group invariants.  This generically requires multiple insertions of $\mathcal{Y}$.  Schematically $y \sim \mathcal{Y}^k$, where $k$ is some integer.  On the other hand, the leading BSM flavour effects are captured by the Wilson coefficients $\mathcal{W}$ of the dimension-6 SMEFT operators which may be in $\mathbf{3}$'s, $\mathbf{6}$'s, $\mathbf{8}$'s, or even $\mathbf{27}$'s of the various SU$(3)$ flavour groups.  The way in which $\mathcal{W}$ can arise depends on the irrep of $\mathcal{Y}$. As such the patterns of observable flavour violation can deviate significantly from expectations based on traditional Ans\"{a}tze.  As we will see, it is possible to have SMEFT flavour violation which is non-hierarchical, or even enhanced, originating from the same spurion which yields a hierarchical Yukawa.

The flavour hierarchies can arise entirely \emph{by accident}, from spurions with a handful of untuned $\mathcal{O}(1)$ elements. For some instances of $\mathcal{Y}$ the leading Yukawa matrix that can be generated is rank-1, thus giving rise to only one non-zero eigenvalue.  Generating a higher-rank $y$ requires more insertions of $\mathcal{Y}$.  This corresponds to a higher dimension operator in the IR theory however and is thus accompanied by some additional suppression factor, which we will denote $\epsilon$. It transpires that there are patterns of higher irrep $\mathcal{Y}$, with generic, untuned entries, which generate a Yukawa matrix $y$ which is rank-1 at leading order, rank-2 at $\mathcal{O}(\epsilon)$, and rank-3 only at $\mathcal{O}(\epsilon^2)$. This is similar to a mechanism of radiative generation of the flavour hierarchies (see, e.g., \cite{Weinberg:1972ws,Georgi:1972mc,Cheung:2010hk}); however, because the spurion is in a non-minimal irrep, the pattern of symmetry breaking is different. We thus naturally, accidentally, generate a hierarchical SM flavour structure from a vastly different structure in the UV.

In a parallel vein, for a scenario wherein $y \sim \mathcal{Y}^k$ one may explain an enhanced flavour hierarchy arising from a smaller one in $\mathcal{Y}$. Essentially the non-Abelian nature gives rise to a \emph{hierarchy multiplier}, with implications for model-building and our picture of UV flavour.

\section{Models}
\label{sec:models}
Here we wish to understand the form of the Yukawa couplings $[y_{q}]^a_\alpha \overline{Q}_{a} H q^\alpha$ where $q \in \{u,d\}$; we discuss the extension to the lepton sector in \cref{sec:leptons}. We may think of this operator as an invariant under $\text{SU}(3)_Q \times \text{SU}(3)_{q}$. How can it arise from a spurion that is not in a $\mathbf{3}\times \overline{\mathbf{3}}$?

We consider a spurion $\mathcal{Y}$ in a $\mathbf{6} \times \overline{\mathbf{6}}$.\footnote{We adopt the convention in which the $\mathbf{6}$ has Dynkin labels $(0,2)$.  Also note that smaller, asymmetrical options for the spurion irrep are $\mathbf{3} \times \overline{\mathbf{6}}$ and $\mathbf{6} \times \overline{\mathbf{3}}$.} We represent this spurion as $Y_{(ab)}^{(\alpha\beta)}$, built from (the symmetric combination of) two anti-fundamental indices $a,b$ of $\mathrm{SU}(3)_Q$, and two fundamental indices $\alpha,\beta$ of $\text{SU}(3)_{q}$. In this case the lowest-order contributions to the Yukawa are of the form
\begin{equation}\label{eq:expansion}
y  \sim  \mathcal{Y}^2 \overline{\mathcal{Y}} + \epsilon \left(\mathcal{Y}^4  + \mathcal{Y} \overline{\mathcal{Y}}^3 \right) +  \epsilon^2 \left( \overline{\mathcal{Y}}^5 + \mathcal{Y}^3 \overline{\mathcal{Y}}^2 \right)  + \ldots \, ,
\end{equation}
where the explicit tensor contractions are given in \cref{eq:explicitContractions}, and we assume that additional insertions of $\mathcal{Y}$ are suppressed by $\epsilon \lesssim 1$.
If we wish to explain the magnitude of the top quark Yukawa coupling, the $\mathcal{O}(\mathcal{Y}^3)$ term ought to be $\mathcal{O}(1)$.  This creates some tension in the IR power counting ---  we comment on the UV implications of this in App.~\ref{app:power}. Note that the power counting of \cref{eq:expansion} is inherently radiatively stable against self-corrections to $\mathcal{Y}$ of the form $\delta Y = \epsilon^\prime \overline{\mathcal{Y}}^2$, for $\epsilon^\prime \lesssim \epsilon$.

We will now consider two novel ways in which a flavour hierarchy may be explained by a non-minimal spurion.

\subsection{Flavour by Accident\label{sec:flavourByAccident}}

Consider a $\mathcal{Y}$ in which the only independent non-zero components are as listed in Tab.~\ref{tab:model}, alongside their charges under the four different Abelian subgroups of $\text{SU}(3)_Q \times \text{SU}(3)_{q}$, in the convention where the corresponding generators in the fundamental irrep are $\mathrm{diag}(1,-1,0)$ and $\mathrm{diag}(1,1,-2)$. We assume the spurion elements to be $\mathcal{O}(1)$, but we need not assign them any specific values.  With such a symmetry-breaking spurion the Yukawa matrix is automatically of the form
\begin{equation}
y \approx
\begin{pmatrix}
    \mathcal{O}(\epsilon) & \mathcal{O}(\epsilon) & 0 \\
    0 & 0 & \mathcal{O}(\epsilon)  \\
    0 & \mathcal{O}(\epsilon) & \mathcal{O}(1)
\end{pmatrix} + \mathcal{O}(\epsilon^2)~~,
\label{eq:Yuk}
\end{equation}
at leading orders in the counting of Eq.~\ref{eq:expansion}.  Performing the singular value decomposition of this matrix, one finds quark masses of $(\mathcal{O}(\epsilon^2),\mathcal{O}(\epsilon),\mathcal{O}(1))$, precisely as desired for an accidental explanation of the flavour hierarchies. For the up (down) sector, we require $\epsilon \equiv \epsilon_u \approx 0.003$ ($\epsilon \equiv \epsilon_d \approx 0.03$).

\begin{table}[t]
\caption{Non-zero spurion elements and U$(1)$ charges in the model of \cref{sec:flavourByAccident}.}
\centering
\begin{tabular}{c c c c c}
\toprule
Element & $Q_3$ & $Q_8$ & $q_3$ & $q_8$ \\\midrule 
$\mathcal{Y}_{12}^{11}$ & 0 &-2&2&2 \\
$\mathcal{Y}_{13}^{23}$ & -1 &1 &-1&-1 \\
$\mathcal{Y}_{22}^{13}$ & 2 &-2&1&-1 \\
$\mathcal{Y}_{22}^{23}$ & 2 &-2&-1&-1 \\\bottomrule
\end{tabular}
\label{tab:model}
\end{table}
One may understand how this construction is distinct from FN models by symmetry considerations alone. 
Considering the fermion charges under all $\mathrm{U}(1)$ subgroups, as given in \cref{tab:model}, we see that there are additional terms which would spoil the predicted mass hierarchy  which are allowed by the Abelian symmetries, but are absent due to their non-Abelian charges. These additional undesirable contributions, which all arise at cubic order in the spurion, are detailed in Tab.~\ref{tab:elements}. We thus see that the generated hierarchy cannot result from Abelian symmetries alone, and is instead a consequence of the genuinely non-Abelian nature of the model.

\begin{table}
\caption{Cubic contributions to the Yukawa allowed by the $\mathrm{U}(1)$ charges of the spurion elements in Table~\ref{tab:model}, but forbidden by their non-Abelian charges. If these were present it would spoil the mass hierarchy.}
\centering
\begin{tabular}{c  c}
\toprule
Entry & Combination \\\midrule
$y_{21}$ & $\mathcal{Y}_{12}^{11} \mathcal{Y}_{13}^{23} \overline{\mathcal{Y}}^{12}_{11}$,
$\mathcal{Y}_{13}^{23} \mathcal{Y}_{13}^{23} \overline{\mathcal{Y}}^{13}_{23}$,
$\mathcal{Y}_{13}^{23} \mathcal{Y}_{22}^{13} \overline{\mathcal{Y}}^{22}_{13}$,
$\mathcal{Y}_{13}^{23} \mathcal{Y}_{22}^{23} \overline{\mathcal{Y}}^{22}_{23}$  \\
$y_{22}$ & $\mathcal{Y}_{13}^{23} \mathcal{Y}_{22}^{13} \overline{\mathcal{Y}}^{22}_{23}$  \\\bottomrule
\end{tabular}
\label{tab:elements}
\end{table}

We find that a minimum of four elements in $\mathcal{Y}$ must be non-zero in order to reproduce the SM Yukawa texture. Whilst there are numerous alternative combinations of four or more elements that can also produce the intended structure, we choose to present the elements in Tab.~\ref{tab:model} as all three quark masses are generated at $\mathcal{O}(\mathcal{Y}^4)$, due to a seesaw mechanism. We therefore find this to be the most minimal and intuitive example. Note however, as shown in Appendix \ref{app:potential}, such a spurion could not arise as a result of spontaneous flavour symmetry breaking by a scalar field with a renormalizable potential. More generally, the possible symmetry breaking patterns of the renormalizable potential of any scalar field do not generate a $(\mathcal{O}(\epsilon^2),\mathcal{O}(\epsilon),\mathcal{O}(1))$ hierarchy in fermion masses.  As a result, if the spurion values are assumed to arise dynamically, which is not a necessity, then they ought to arise from a non-renormalizable scalar potential or possibly non-perturbative dynamics.

\subsection{Hierarchy Multiplier\label{sec:hierarchyMultiplier}}

Another notable property of $\mathcal{Y} \sim \mathbf{6} \times \overline{\mathbf{6}}$ is that the lowest order at which a Yukawa can arise is cubic.  As a result, if some elements of $\mathcal{Y}$ are $\mathcal{O}(1)$, and others $\mathcal{O}(\delta)$, then the elements of the Yukawa could range from $\mathcal{O}(1)$ to $\mathcal{O}(\delta^3)$. In other words, due to the representation itself, mild hierarchies in flavour in the UV could be enhanced when propagating down to the IR.

Realizing this in practice is complicated, because the presence of Levi-Civita epsilons in the appropriate contraction of $\mathcal{Y}^2 \overline{\mathcal{Y}}$ mean that elements of the Yukawa are products of different elements of $\mathcal{Y}$, as opposed to the same element cubed. This can impede the `multiplication' effect. For example, if the elements of $\mathcal{Y}$ are drawn from a log-uniform prior over the range $[\delta,1]$, and then $y \sim \mathcal{Y}^2 \overline{\mathcal{Y}}$ constructed, the resulting first generation mass is larger on average than if the elements of $y$ were sampled directly from the same distribution.\footnote{This choice of prior is purely illustrative: it is notably not an invariant measure on $\mathrm{SU}(3) \times \mathrm{SU}(3)$.} For a hierarchy within $\mathcal{Y}$ to be amplified, the elements need to have a non-trivial structure.


\begin{table}[t]
\caption{Non-zero spurion elements in the model of \cref{sec:hierarchyMultiplier}}
\centering
\begin{tabular}{l@{\hskip 10pt} c c c c c }\toprule
Element &
$\mathcal{Y}_{33}^{33}$ &
$\mathcal{Y}_{13}^{13}$ &
$\mathcal{Y}_{22}^{23}$ & 
$\mathcal{Y}_{23}^{22}$ & 
$\mathcal{Y}_{12}^{12}$ \\[1ex]
Scaling 
& $\mathcal{O}(\delta)$  
& $\mathcal{O}(\delta)$ 
& $\mathcal{O}(1)$  
& $\mathcal{O}(1)$ 
& $\mathcal{O}(1)$ \\\bottomrule
\end{tabular}
\label{tab:modelmult}
\end{table}

We may illustrate the mechanism through a simple example.  Consider a spurion with non-zero elements as detailed in Tab.~\ref{tab:modelmult}, which scale such that those that couple to the third generation in both of the two flavour groups are associated with a factor of $\delta$. At leading cubic order this gives rise to a mass matrix scaling as
\begin{equation}
y \approx 
\begin{pmatrix}
    \mathcal{O}(\delta^2) & 0 & 0 \\
    0 & \mathcal{O}(\delta) & 0  \\
    0 & 0 & \mathcal{O}(1)
\end{pmatrix} ~~,
\label{eq:Yukmult}
\end{equation}
where we see that a spurion whose smallest element is $\mathcal{O}(\delta)$ gives rise to a Yukawa whose smallest eigenvalue scales as $\mathcal{O}(\delta^2)$.  Whether this scaling is preserved at quartic and higher orders in the spurion expansion \cref{eq:expansion} depends on the underlying power counting of the model, e.g.\ if there is $\mathbb{Z}_2$ symmetry of the spurion.  Furthermore, examples with more extreme scaling at leading order are possible.  However, the above case illustrates a complementary mechanism to \cref{sec:flavourByAccident}, through which the breaking of flavour symmetries by higher order irreps enhances hierarchies in the SM Yukawas. We do not consider this mechanism further in this paper.

\section{Phenomenology}
\label{sec:pheno}

In MFV models one generically expects the pattern of dimension-6 flavour violation to follow that of the Yukawa hierarchies, or powers thereof.  In FN models this aspect is more model-dependent (see e.g.\ \cite{Lalak:2010bk,Antusch:2023shi,Greljo:2024zrj}).  In the models introduced in \cref{sec:models}, whilst the leading order contribution to the Yukawa arises at cubic order in the spurion, the flavour-violating dimension-6 SMEFT operators can be generated from just 2 spurion insertions. In this way, the flavour violation in the SMEFT is correlated with the Yukawa hierarchies differently from traditional scenarios. 

To illustrate this we return to the `accidental hierarchy' example spurion defined in Table~\ref{tab:model}. The flavour violating (i.e.\ non-singlet) irreps that can appear in the decomposition of the Wilson coefficients of the dimension-6 four-fermion SMEFT operators include the $\mathbf{8}$s and $\mathbf{27}$s of both $\mathrm{SU}(3)_Q$ and $\mathrm{SU}(3)_q$ in addition to the $\mathbf{8} \times \mathbf{8}$s of $\mathrm{SU}(3)_Q \times \mathrm{SU}(3)_q$.

One can characterise the pattern of flavour violation in these irreps via the leading order contribution in $\epsilon$ to each component, when these irreps are formed from the appropriate contractions of $\mathcal{Y}$ and rotated to the mass basis, i.e.\ the up(down) basis for $q=u(d)$. We find that the coefficients of operators built from left- and right-handed fields scale as
\begin{subequations}
\begin{align}
    \left(\bar{Q} \gamma Q\right)^2: \quad & c_{LL} \approx \epsilon^\frac{\abs{Q_8 - Q_3}}{2} \, ,\\
    \left(\bar{q} \gamma q\right)^2: \quad & c_{RR} \approx \epsilon^\frac{\abs{q_8}}{3} \, ,\\
    \left(\bar{Q} \gamma Q\right) \left(\bar{q} \gamma q\right): \quad & c_{LR} \approx \epsilon^{\frac16 \abs{3 Q_3 - 2 q_8 - Q_8} + (Q_3 \, \mathrm{mod} \, 2)} \, .
\end{align}%
\label{eq:AHSpurionPattern}%
\end{subequations}%
We have written the above in terms of the Abelian charges of the components. All components with non-zero charges have some degree of off-diagonality; only those with non-zero $Q_8$ or $q_8$ involve the third generation. Each chirality of four-quark operator has a suppressed flavour-violating direction in the space of Abelian charges, but `orthogonal' directions are unsuppressed. Contrast to the case of MFV, wherein all flavour-violating directions are to some extent suppressed \cite{DAmbrosio:2002vsn,Cirigliano:2006su}.

The predicted pattern of FN models is closer in spirit but still different. FN models also produce a texture of suppressed and unsuppressed flavour violating directions scaling as
\begin{equation}
    \dot\epsilon^{\abs{a Q_3 + b Q_8 + c q_3 + d q_8}} 
\end{equation}
in the dimensionless vev of the flavon field $\dot\epsilon$, for some $a,\ldots,d$ that are linear combinations of the U$(1)_\text{FN}$ charges of the quarks. This can almost fit the pattern \cref{eq:AHSpurionPattern}, with the exception of some elements of $c_{LR}$. More importantly, the U$(1)_\text{FN}$ charges required to approximate \cref{eq:AHSpurionPattern}, when $\epsilon=\dot\epsilon$, predict significantly different masses for the down quarks of $\mathcal{O}(1),\mathcal{O}(\dot\epsilon),\mathcal{O}(1)$. In other words, the pattern in the dim-6 coefficients is similar but not identical; however, the correlation between dim-6 and dim-4 coefficients differs vastly for the two models.

To emphasize the different phenomenology that the accidental hierarchy model can lead to, in \cref{tab:mixing} we list the size of the $\Delta F_d = 2$ components of the four-quark operators which mediate $K^0$, $B^0$, and $B_s$ oscillations. In MFV, transitions between lighter flavours are suppressed by light Yukawas, and in the down basis the right-handed flavour violating operators are further suppressed by down-type Yukawa insertions. These trends are precisely the opposite for the example spurion given in \cref{tab:model}, to the extent that it mediates $O(1)$ $K^0$--$\overline{K}^0$ oscillations via the right-handed operator $\left(\bar{d}_R \gamma s_R\right)^2$. Clearly, this prediction of unsuppressed mixing between light fermions differs starkly from the traditional MFV paradigm. A viable FN model necessarily has different directions of unsuppressed mixing if it reproduces the down quark masses: in $c_{LR}$ for $sd$ and in $c_{RR}$ for $bs$.

Bounds on Kaon mixing push the permitted scale of flavour breaking in this accidental hierarchy model significantly above the TeV scale. The bounds on the real coefficient of the right-handed operator $\left(\bar{d}_R \gamma s_R\right)^2$, $\mathrm{Re} \, c_{RR} \lesssim 6.5 \times 10^{-7} \, \mathrm{TeV}^{-2}$ \cite{Silvestrini:2018dos}, suggest a scale of $(\mathrm{Re} \, c_{RR})^{-\frac12} \approx 10^3 \, \mathrm{TeV}$, in light of the lack of suppression of the Wilson coefficient. There is a tension between the flavour and electroweak scales.

The above is an instance of one spurion in one particular irrep; others give different flavour structures in SMEFT operators.  For instance, there are alternative spurions which lead to suppressed Kaon mixing.\footnote{E.g, a four-entry spurion $\widetilde{\mathcal{Y}}$ with non-zero elements $\left\{\widetilde{\mathcal{Y}}_{11}^{12},\widetilde{\mathcal{Y}}_{11}^{13},\widetilde{\mathcal{Y}}_{12}^{13},\widetilde{\mathcal{Y}}_{23}^{22}\right\}$ gives rise to suppressed flavour violation in the light generations with a no-greater-than-$\mathcal{O}(\epsilon^2)$ coefficient.}  Generally speaking, we have demonstrated here a different correlation between SM and NP flavour structures when they are generated by the same flavour breaking spurion. In particular, the effects of NP from a single non-Abelian spurion are not necessarily enhanced in the third generation --- that is specific to MFV, when the spurion is in the $\mathbf{3} \times \overline{\mathbf{3}}$ irrep.

\begin{table}
\caption{In the down basis, the suppression of $\Delta F_d = 2$ flavour-violating processes relative to flavour-conserving ones, as mediated by new physics in an MFV scenario, the `accidental hierarchy' scenario (AH) defined in \cref{sec:flavourByAccident}, and a Froggatt-Nielsen scenario (FN). $s_{ij} \equiv \sin \theta_{ij}$ are the CKM mixing angles in the convention of \cite{Chau:1984fp}. $\epsilon = \epsilon_d \approx 0.03$, although there are also subdominant contributions to $c_{LL}$ built from equal powers of $\epsilon_u$. $\dot\epsilon \approx 0.2$, and we assume the U$(1)_\text{FN}$ charges of the model of \cite{Leurer:1993gy}, which predicts down quark masses of $\mathcal{O}(\dot\epsilon^7),\mathcal{O}(\dot\epsilon^5),\mathcal{O}(\dot\epsilon^3)$. For other FN models see \cite{Cornella:2023zme}.\label{tab:mixing}}
\vspace{1ex}
\begin{tabular}{@{}lccc}\toprule
    & MFV & AH & FN\\\midrule
    & \multicolumn{3}{c}{$sd$}\\\cmidrule(l){2-4}
    $c_{LL}$ & $y_c^4 s^2_{12}$ & $\epsilon^2$ & $\dot\epsilon^2$\\
    $c_{LR}$ & $y_c^4 s^2_{12} y_d y_s$ & $\epsilon^1$ & $1$ \\
    $c_{RR}$ & $y_c^4 s^2_{12} y_d^2 y_s^2$  & $1$ & $\dot\epsilon^2$ \\\midrule
    & \multicolumn{3}{c}{$bd$}\\\cmidrule(l){2-4}
    $c_{LL}$ & $y_t^4 s^2_{12} s^2_{23}$ & $\epsilon^2$ & $\dot\epsilon^6$ \\
    $c_{LR}$ & $y_t^4 s^2_{12} s^2_{23} y_d y_b$ & $\epsilon^2$ & $\dot\epsilon^2$\\
    $c_{RR}$ & $y_t^4 s^2_{12} s^2_{23} y_d^2 y_b^2$ & $\epsilon^2$ & $\dot\epsilon^2$\\\midrule
    & \multicolumn{3}{c}{$bs$}\\\cmidrule(l){2-4}
    $c_{LL}$ & $y_t^4 s^2_{23}$ & $\epsilon^4$ & $\dot\epsilon^4$ \\
    $c_{LR}$ & $y_t^4 s^2_{23} y_s y_b$ & $\epsilon^3$ & $\dot\epsilon^2$\\
    $c_{RR}$ & $y_t^4 s^2_{23} y_s^2 y_b^2$ & $\epsilon^2$ & $1$ \\\bottomrule
\end{tabular}
\end{table}

\section{CKM}
This structure may be straightforwardly extended to accommodate all charged fermions of the SM, including the generation of the CKM matrix.  Consider $\mathcal{Y}_u$ as a $\mathbf{6} \times \overline{\mathbf{6}}$ under $\text{SU}(3)_Q \times \text{SU}(3)_{u}$ and $\mathcal{Y}_d$ as a $\mathbf{6} \times \overline{\mathbf{6}}$ under $\text{SU}(3)_Q \times \text{SU}(3)_{d}$.  The Yukawa matrices $y_u,y_d$ constructed out of these spurions transform as a $\mathbf{3} \times \overline{\mathbf{3}}$ under $\text{SU}(3)_Q \times \text{SU}(3)_{u}$ and $\text{SU}(3)_Q \times \text{SU}(3)_{d}$ respectively.

If one were to take $\mathcal{Y}_u = \mathcal{Y}_d$ then the Yukawa matrices would be identical.  However, since the group transformations are well-defined, if we take instead that $\mathcal{Y}_u = G_Q \left(\mathcal{Y}_d \right)$, where the latter represents an $\text{SU}(3)_Q$ rotation of the spurion, then we have that $y_u = V y_d$ where $V$ is a unitary matrix.  In this way, choosing $V$ to be the CKM matrix straightforwardly accommodates the desired SM structure.  The difference in Yukawa eigenvalues can follow from the freedom of the expansion terms in Eq.~\ref{eq:expansion}, which allows for different expansion parameters for the different flavours, $\epsilon_u$ and $\epsilon_d$.

We may go further and ask if our choice of spurion can naturally lead to a hierarchy in the CKM. This may provide a possible explanation for the values of the CKM within in our framework rather than merely accommodating them. Within the accidental hierarchy model of \cref{sec:flavourByAccident}, we construct the Yukawas from similar spurions $\mathcal{Y}_u$ and $\mathcal{Y}_d$, such that both $y_u$ and $y_d$ have the same texture as Eq.~\ref{eq:Yuk} but with different $\mathcal{O}(1)$ values. We extract the left-handed unitary matrices required to diagonalize the Yukawas
\begin{equation}
U_q^L =
\begin{pmatrix}
    \mathcal{O}(\epsilon_q) & \mathcal{O}(1) & \mathcal{O}(\epsilon_q^2) \\
    \mathcal{O}(1) & \mathcal{O}(\epsilon_q) & \mathcal{O}(\epsilon_q)  \\
    \mathcal{O}(\epsilon_q) & \mathcal{O}(\epsilon_q^2) & \mathcal{O}(1)   
\end{pmatrix}
\end{equation}
and find a CKM of the form
\begin{equation}
V_{\text{CKM}} = U_u^L (U_d^L)^\dagger = 
\begin{pmatrix}
    \mathcal{O}(1) & \mathcal{O}(\epsilon) & \mathcal{O}(\epsilon) \\
    \mathcal{O}(\epsilon) & \mathcal{O}(1) & \mathcal{O}(\epsilon^2)  \\
    \mathcal{O}(\epsilon) & \mathcal{O}(\epsilon^2) & \mathcal{O}(1)
\end{pmatrix} \, ,
\label{eq:CKM}
\end{equation}
where each $\epsilon$ can be either $\epsilon_u$ or $\epsilon_d$. While this particular spurion does not reproduce the measured pattern of CKM values; it is nonetheless encouraging that the `accidental hierarchy' setup can generate a hierarchical CKM at the same time as hierarchical mass. It is plausible that with a different choice of spurion entries or irrep the desired pattern could arise. Moreover, CP violation is naturally small: if the entries of $\mathcal{Y}_u$ and $\mathcal{Y}_d$ are given $\mathcal{O}(1)$ phases, then due to the small CKM mixing the Jarlskog invariant is the correct order of magnitude.

\section{Leptonic Textures\label{sec:leptons}}

We note briefly that this spurion analysis can be extended to the leptonic sector. We start with the Lagrangian
\begin{equation}
     \lag = - [y_e]^a_\alpha \overline{L}_a H e^\alpha + c_{ab} (HL^a) (HL^b) + \text{h.c.}  \, , 
\end{equation}
where $y_e \sim \mathbf{3} \times \overline{\mathbf{3}}$ is a hierarchical Yukawa transforming under $\mathrm{SU}(3)_L \times \mathrm{SU}(3)_e$ and $c \sim \mathbf{6} \times \mathbf{1}$ describes Majorana neutrino masses. $c$ should not be aligned with the Yukawa.

Much like the up and down Yukawas in the quark sector, we cannot generate both $y_e$ and $c$ from multiple insertions of a single spurion, and require at least two.\footnote{This follows from the irreps' triality, as defined in Appendix \ref{app:reptheory}. Since the Yukawa has $(T_L,T_e) = (1,2)$, this would require that a single spurion irrep has both $T_L$ and $T_e$ non-zero. This then makes it impossible to produce $c$, which has $(T_L,T_e)=(1,0)$.} Considering the minimal case of two spurions, the hierarchical $y_e \sim \mathcal{Y}_e^2 \overline{\mathcal{Y}}_e$ could, for example, be built from a $\mathcal{Y}_e \sim \mathbf{6} \times \overline{\mathbf{6}}$, a leptonic analogue of the quark constructions in \cref{sec:models}. The desired structure of $c$ is different. It is preferable to have many insertions of a spurion to suppress the Weinberg operator in the IR as much as possible.

A few small irreps (such as the $\mathbf{21}$) first produce a $\mathbf{6}$ at quartic order. If the second spurion $\mathcal{S} \sim \mathbf{1} \times \mathbf{21}$ then
\begin{equation}
    c \sim \frac{\mathcal{Y}_e \mathcal{S}^4}{\Lambda^6} \, ,
\end{equation}
and the neutrino mass $m_\nu \lesssim 0.1 \, \mathrm{eV}$ then sets the scale to be
\begin{equation}
    \Lambda \approx \frac{v^2}{m_\nu} \left( \frac{\abs{\left\langle \mathcal{Y}_e \right\rangle}}{\Lambda} \right) \left( \frac{\abs{\left\langle \mathcal{S} \right\rangle}}{\Lambda} \right)^4 \, .
\end{equation}
If there is a moderate hierarchy $\abs{\left\langle \mathcal{Y}_e \right\rangle},\abs{\left\langle \mathcal{S} \right\rangle} \approx 0.01 \Lambda$, then the scale of flavour breaking $\Lambda$ is in the TeV range. The left-handed rotations required to diagonalize $y_e$ and $c$ will generally differ, accommodating $\mathcal{O}(1)$ mixing in the PMNS matrix.

\section{Conclusions}
\label{sec:conclusions}

Flavour has played a crucial role in developing our understanding of the SM, yet much remains unanswered. The hierarchical  patterns adorning the  flavour sector persist as one of the greatest structural enigmas of the SM. With a substantial ongoing experimental programme probing flavour violation across a variety of phenomena to ever-increasing precision, it is timely to re-evaluate our theoretical expectations as to what we may hope to learn from these efforts. 

From a theoretical  standpoint the outcome can be understood in terms of the pattern of flavour violation  in the SMEFT operators, which emerges from the way in which the flavour symmetries are broken in the deep UV. To what extent have we thus far explored the landscape of possibilities for the UV? One concrete way to address this is to consider how the flavour violating parameters of our theories transform under flavour symmetries. 

In this work we have demonstrated that the traditional narratives are just one possible story that may play out in the UV. In particular, we highlight that there is an entire, heretofore unexplored, class of models in which the parameters responsible for breaking the flavour symmetry of the gauge sector transform in non-minimal irreducible representations of this group. We have shown that in such models it is possible to multiply or accidentally generate the hierarchies in fermion masses by virtue of the non-Abelian nature of the flavour symmetries. In addition, we have demonstrated that the pattern of flavour violation appearing in the dim-6 SMEFT operators can be correlated with the Yukawa hierarchies in unexpected ways, allowing for strikingly different phenomenology to the traditional approaches, including (but not limited to) the possibility of  unsuppressed flavour violation amongst the light quarks. This result strengthens the case for a broad experimental flavour programme. 

We note another rationale for considering non-minimal irreps, other than the obvious desire to map out the full landscape of possibilities for the UV and their resulting phenomenology. The dimension-5 Weinberg operator is itself in the $\mathbf{6}$ of $\mathrm{SU}(3)_L$. Should therefore neutrinos be Majorana in nature, the flavour symmetries would necessarily be broken non-minimally. Non-minimal flavour symmetry breaking may already play a role in nature.

\acknowledgments{The authors are grateful to Admir Greljo, Gino Isidori and Jure Zupan for valuable comments on an early draft.  M.M.\ would also like to thank Neal Weiner for discussions. This research was supported in part by grant NSF PHY-2309135 to the Kavli Institute for Theoretical Physics (KITP). H.B.\ acknowledges partial support from the STFC HEP Theory Consolidated grants ST/T000694/1 and ST/X000664/1 and thanks other members of the Cambridge Pheno Working Group for useful discussions. G.C.\ is supported by STFC grant ST/Y509188/1. D.S.\ is supported by STFC grant ST/X000605/1.}

\appendix

\section{Comments on Power Counting}
\label{app:power}
The power counting for the `accidental hierarchy' scenario requires that higher order terms in the EFT expansion of the fermion Yukawa, Eq.~\ref{eq:expansion}, are suppressed relative to the leading $\mathcal{O}(\mathcal{Y}^3)$ term.  On the other hand, the leading fermion mass arises already at $\mathcal{O}(\mathcal{Y}^3)$.  For the down-type quarks and charged leptons these two requirements aren't obviously in significant tension.  However, for the up-type quarks they are, since one requires that, in natural units, the top quark Yukawa is $\mathcal{O}(1)$.  This question concerns the QFT power counting and hence the overall structure of the UV scenario.  Addressing it is thus mostly beyond the scope of this work.  Nonetheless, here we sketch a few possible UV scenarios, whilst reminding the reader that this aspect remains an open question to be resolved.

\subsection*{Special Top}
One possibility is that the top Yukawa does not fit within this framework and in the up sector one has only a U$(2)$ symmetry, as opposed to the rest of the fermions.  In the language of the SU$(3)$ symmetries this would require that one has a single spurion $\mathcal{Y}_T \sim \mathbf{3} \times \overline{\mathbf{3}}$ giving rise to a rank-1 Yukawa for the top mass.  In addition to this, one would have the usual $\mathcal{Y} \sim \mathbf{6} \times \overline{\mathbf{6}}$ which would now break the remaining symmetries for the up and charm masses.  Potentially significant cross-terms between the two spurions could be evaded if, for instance, $\mathcal{Y}_T$ was only non-vanishing in the $33$ component, whereas $\mathcal{Y}$ is only non-vanishing in all $1,2$ components.  This is a rather cludgy explanation for the structure of the up-quark masses, but a logical possibility nonetheless.

\subsection*{Dynamical Origins}
We now sketch an example UV scenario to motivate how the required power counting may arise in the UV.  We treat $\mathcal{Y}$ as a spurion in the $\mathbf{6} \times \overline{\mathbf{6}}$ of the flavour symmetry, whose entries are the only source of explicit symmetry breaking.

\newcommand{\fh}{F}
To generate the up Yukawa, we also introduce a Higgs-like scalar $\fh$, in a $\mathbf{2}_{-\frac12}$ of $\mathrm{SU}(2)_L \times \mathrm{U}(1)_Y$ and a $\mathbf{3}_Q  \times \overline{\mathbf{3}}_u$ of the flavour symmetries, as well as a neutral scalar $\mathcal{X}$ in the $\mathbf{3}_Q \times \overline{\mathbf{3}}_u$ irrep, and assume that neither acquire a symmetry-breaking vev.

The renormalizable interactions allowed under the SM gauge symmetries and flavour symmetry are
\begin{align}
& \tilde{y} \overline{Q} \fh  u + \lambda \fh H \mathcal{X}^2 + \lambda' \fh H \mathcal{X} \mathcal{Y} + \mu \overline{H} \overline{\fh} \mathcal{X} \notag\\
&+  A \mathcal{X}^3  + A' \mathcal{Y}^3 + A'' \mathcal{X}^2\mathcal{Y} +  B \overline{\mathcal{X}}^2\mathcal{Y}^2 + B' \mathcal{X}^2 \overline{\mathcal{X}} \overline{\mathcal{Y}} \notag\\
&+  B'' \mathcal{X}\mathcal{Y}\overline{\mathcal{Y}}^2 + C \fh \overline{\fh} \mathcal{X}\overline{\mathcal{Y}} + \textnormal{h.c.} + \textnormal{cross quartics} \, , \label{eq:scalarPot}
\end{align}
where cross quartics refers to all terms of the form $\lambda_\times \, \mathcal{V}\overline{\mathcal{V}} \mathcal{W}\overline{\mathcal{W}}$ where $\mathcal{V}, \mathcal{W} \in \{\mathcal{X},\mathcal{Y},\fh \}$ and $\lambda_\times$ is some coupling. Integrating out $\mathcal{X}$ and $\fh$ one generates
\begin{eqnarray}
 y & = & \tilde{y} \frac{(B'')^* \mu}{M^2_{\mathcal{X}} M^2_{\fh}} \mathcal{Y}^2 \overline{\mathcal{Y}} \, ,  \\
y_\epsilon & = & \tilde{y} \frac{B'' (\lambda')^*}{M^2_{\mathcal{X}} M^2_{\fh}}  \mathcal{Y}\overline{\mathcal{Y}}^3 \, ,
\end{eqnarray}
for the pieces of the Yukawa that are cubic and quartic in $\mathcal{Y}$ respectively. Alternatively, to generate the down Yukawa, the hypercharge of $\fh$ must be inverted, and $H$ insertions in Lagrangian \cref{eq:scalarPot} conjugated.

Setting $M_\mathcal{X} \approx M_{\fh} \approx \abs{\mu} \approx M$ and $\abs{B''} \approx \abs{\tilde{y}}^2 \approx (4\pi)^2$, a small hierarchy between the vev and mass scales of $\abs{\left\langle \mathcal{Y}\right\rangle} \approx 0.1 M$ gives an order-one top Yukawa $y$. Relative to the top Yukawa, the following term is suppressed
\begin{equation}
|y_{\epsilon}| \approx |y| \times \frac{  \abs{\langle \mathcal{Y}\rangle} \abs{\lambda'}}{ \abs{\mu}} \, .
\end{equation}
This can naturally be small if, for example, all fields charged under $\mathrm{SU}(3)_Q$ are also odd under a $\mathbb{Z}_2$ symmetry, which is broken by $\lambda'$.

The quintic terms in $\mathcal{Y}$ can only be generated at tree level by an insertion of a cross quartic $\lambda_\times$ into a diagram that generates the leading Yukawa term:
\begin{equation}
y_{\epsilon^2} = \left( \frac{\lambda_\times}{M^2_{\mathcal{X}}} \text{ OR } \frac{\lambda_\times}{M^2_{\fh}} \right) \mathcal{Y}\overline{\mathcal{Y}} \times  y \, .
\end{equation}
While this may not be sufficiently suppressed, the cross quartics generate a limited number of possible contractions of the flavour indices of the five $\mathcal{Y}$ insertions. For the example spurion given in \cref{tab:model}, all such contractions are zero, meaning that the quintic terms do not spoil the hierarchy amongst Yukawa elements.

\subsection*{Extra Symmetries}
Another possibility to reconcile the magnitude of the leading and subleading terms in \cref{eq:expansion} is if there are additional discrete symmetries at play.  For instance, suppose $\mathcal{Y}$ were subject to an additional $\mathbb{Z}_2$ selection rule, arising from the UV.  In this case Yukawa contributions would only arise at $\mathcal{O}(\mathcal{Y}^3,\mathcal{Y}^5,\mathcal{Y}^7,...)$.  As a result there would be more freedom to create large hierarchies between subsequent terms, despite the $\mathcal{O}(\mathcal{Y})^3$ term being $\mathcal{O}(1)$, after accounting for additional factors of couplings and so on.  This scenario would then consist of a mix of accidental discrete and continuous symmetries at play.

\subsection*{Alternative Irreps}

The tension in the power counting between obtaining an order-one top mass and suppressing higher-order terms can be alleviated somewhat by placing the spurion $\mathcal{Y}$ in an irrep that produces the Yukawa at quadratic order
\begin{equation}
y  \sim  \overline{\mathcal{Y}}^2 + \epsilon \mathcal{Y}^2 \overline{\mathcal{Y}} +  \epsilon^2 \left( \overline{\mathcal{Y}}^4 + \mathcal{Y} \overline{\mathcal{Y}}^3 \right)  + \ldots \, .
\end{equation}
In this case the leading order (non-trivial) contributions to both the SM Yukawa and SMEFT coefficients arise at quadratic order; we therefore expect the flavour phenomenology to differ from the $\mathbf{3} \times \overline{\mathbf{3}}$ (MFV) and $\mathbf{6} \times \overline{\mathbf{6}}$ cases.

Consider the minimal example of a $\mathcal{Y}$ in a $\mathbf{15}_Q \times \overline{\mathbf{3}}_u$. The UV completion analogous to \cref{eq:scalarPot} has Lagrangian
\begin{align}
& \tilde{y} \overline{Q} \fh  u 
+ \lambda \fh H \mathcal{X}^2 + \lambda' \fh H \mathcal{Y}^2
 + \mu \overline{H} \overline{\fh} \mathcal{X} \notag\\
&+  A \mathcal{X}^3  + A' \mathcal{Y}^3 + A'' \mathcal{X}\mathcal{Y}^2
 +  B \overline{\mathcal{X}}^2\mathcal{Y}^2 + B' \mathcal{X}^2 \overline{\mathcal{X}} \overline{\mathcal{Y}} \notag\\
&+  B'' \mathcal{X}\mathcal{Y}\overline{\mathcal{Y}}^2 
+ C \fh \overline{\fh} \mathcal{X}\overline{\mathcal{Y}} + \textnormal{h.c.} + \textnormal{cross quartics} \, , 
\end{align}
which means that the first two orders of Yukawa terms are
\begin{eqnarray}
 y & = & \tilde{y} \frac{(\lambda')^*}{M^2_{\fh}} \overline{\mathcal{Y}}^2 \, ,  \\
y_\epsilon & = & \tilde{y} \frac{\mu (B'')^*}{M^2_{\mathcal{X}} M^2_{\fh}}  \mathcal{Y}^2\overline{\mathcal{Y}} \, .
\end{eqnarray}
Not only can the first term be order one while tolerating a larger hierarchy between $\abs{\left\langle \mathcal{Y} \right\rangle}$ and the other mass scales, but said hierarchy also naturally suppresses the next-order terms.

\section{Comments on UV completion via spontaneous symmetry breaking}
\label{app:potential}
There are numerous ways in which a global symmetry-breaking spurion can arise, be it through perturbative or non-perturbative dynamics, or as the result of selection rules arising from accidental symmetries which follow from gauge symmetries and irreps.  For the former, one possibility would be that the spurion which explicitly breaks a global symmetry in the IR, is in fact the expectation of a scalar field which spontaneously breaks a gauge symmetry in the UV, with the UV dynamics integrated out.  A notable example are the light quark masses, which explicitly break chiral symmetry at the QCD scale, but arise from spontaneous breaking of the electroweak symmetry in the UV. We explore this spontaneous symmetry breaking possibility here: what patterns in $\left\langle \mathcal{Y} \right\rangle$ arise from minimizing an $\mathrm{SU}(3) \times \mathrm{SU}(3)$ invariant potential built from $\mathcal{Y}$?\footnote{For a study of this question as it relates to a minimal $\mathbf{3} \times \overline{\mathbf{3}}$ spurion, see \cite{Alonso:2013nca}.}

Up to rescalings, the most general bounded-from-below \emph{renormalizable} potential in $\mathcal{Y}$ can be written
\begin{align}
\label{eq:renSSBPotential}
  V &= (I_2 - 1)^2 + \lambda_3\,I_3 + \lambda_3^\ast\,I_3^\ast \notag\\
  &\hspace{5ex} + \lambda_{4b}\,I_{4b} + \lambda_{4c}\,I_{4c} +  \lambda_{4d}\,I_{4d}  \, ,
\end{align}
for some coefficients $\lambda_i$ multiplying the flavour invariants
\begin{align}
    I_2 &= \,  \mathcal{Y}_{cd}^{\gamma\delta}\overline{\mathcal{Y}}^{cd}_{\gamma\delta} \, , \quad
    I_3 = \, \epsilon^{ace} \epsilon_{\alpha\gamma\epsilon} \epsilon^{bdf} \epsilon_{\beta\delta\zeta} \, \mathcal{Y}_{ab}^{\alpha\beta}\mathcal{Y}_{cd}^{\gamma\delta}\mathcal{Y}_{ef}^{\epsilon\zeta} \, ,  \notag\\
    I_{4b} &= \, \mathcal{Y}_{ab}^{\alpha\beta}  \overline{\mathcal{Y}}^{be}_{\beta\epsilon} \mathcal{Y}_{ef}^{\epsilon\varphi}  \overline{\mathcal{Y}}^{af}_{\alpha\varphi} \, , \quad
    I_{4c} = \, \mathcal{Y}_{ab}^{\alpha\beta}  \overline{\mathcal{Y}}^{ab}_{\beta\epsilon} \mathcal{Y}_{ef}^{\epsilon\varphi}  \overline{\mathcal{Y}}^{ef}_{\alpha\varphi} \, , \notag\\
    I_{4d} &= \, \mathcal{Y}_{ab}^{\alpha\beta}  \overline{\mathcal{Y}}^{be}_{\alpha\beta} \mathcal{Y}_{ef}^{\epsilon\varphi}  \overline{\mathcal{Y}}^{af}_{\epsilon\varphi} \, .
\end{align}
We numerically minimize the potential \cref{eq:renSSBPotential}, scanning over $\lambda_i \in [-2, 2]$ and find phases for $\mathcal{Y}$ which lead to three distinct phases in $y$ given in \cref{tab:SSB}. Minimizing the renormalizable potential generates either: (a) no quark masses;\footnote{Notably this is still possible even if $\left\langle \mathcal{Y} \right\rangle \neq 0$ and the symmetry is non-trivially broken in the $\mathbf{6} \times \overline{\mathbf{6}}$ irrep.} (b) equal quark masses, or (c) just a mass for the third generation. These are the same possibilities for the quark masses that arise from constructing and minimizing a renormalizable potential for $y$ directly \cite{Nardi:2011st}. This result is also consistent with Michel's conjecture that a renormalizable potential of any irrep will break the symmetry to its maximal subgroups \cite{Michel:1979vv}.\footnote{Technically $\mathrm{SU}(2)^2 \times \mathrm{U}(1)^2$, rather than $\mathrm{SU}(2)^2 \times \mathrm{U}(1)$, is a maximal subgroup of $\mathrm{SU}(3) \times \mathrm{SU}(3)$, but all vevs must break $\mathrm{U}(1)_{Q_8} \times \mathrm{U}(1)_{q_8}$ to its diagonal subgroup.}

To achieve a realistic mass texture through spontaneous symmetry breaking, more terms need to be added to the potential, either through non-renormalizable operators, or equivalently through couplings to auxiliary fields that also acquire a vev \cite{Espinosa:2012uu}. It remains to be seen if this extension is easier for a potential built from $\mathcal{Y}$ or directly from $y$.

\begin{table}
\caption{Possible Yukawa vev structures $\langle y \rangle$ and corresponding symmetry breaking patterns when minimizing a renormalizable potential for $\mathcal{Y} \sim \mathbf{6} \times \overline{\mathbf{6}}$. Here $v$ is $\mathcal{O}(1)$.}
\centering
\begin{tabular}{c @{\hskip 15pt} c}
\toprule
$\langle y \rangle$ & Little group of $\langle y \rangle$   \\
\midrule
$0$ & $\mathrm{SU}(3) \times \mathrm{SU}(3)$   \\[5pt]
$\mathrm{diag}(\frac{v}{3},\frac{v}{3},\frac{v}{3})$ & $\mathrm{SU}(3)$   \\[5pt]
$\mathrm{diag}(0,0,v)$ & $\mathrm{SU}(2) \, \times \, \mathrm{SU}(2) \, \times \, \mathrm{U}(1)_{Q_8+q_8}$ \\\bottomrule
\end{tabular}
\label{tab:SSB}
\end{table}

\section{Representation theory\label{app:reptheory}}

An irrep $\mathbf{R}$ of $\mathrm{SU}(3)_f$ is identified by two non-negative integer Dynkin labels $(a_1,a_2)$, which are interchanged in the conjugate irrep. Its triality is given by
\begin{equation}
    T_f(\mathbf{R}) = (a_1 - a_2) \, \text{mod} \, 3 \, .
\end{equation}
A useful constraint on the decomposition $\mathbf{R}_a \times \mathbf{R}_b = \sum_i \mathbf{R}_i$ is \cite{Slansky:1981yr}
\begin{equation}
    T_f(\mathbf{R}_i) = \left( T_f(\mathbf{R}_a) + T_f(\mathbf{R}_b) \right) \, \text{mod} \, 3 \,.
\end{equation}

Restricting w.l.o.g.\ to triality-1 irreps, only those with $(a_1,a_2)=(n+1,n), \,  \forall n \in \mathbb{Z}$, namely the $\mathbf{3},\mathbf{15},\mathbf{42},\ldots$, can produce an antitriplet at quadratic order. Most remaining irreps produce a triplet first at cubic order, except $\mathbf{15^\prime},\mathbf{21},\mathbf{36},\mathbf{45},\mathbf{99}$, and others of higher dimension.

Explicitly, the independent cubic and quartic contractions that form $y^x_\chi$ in \cref{eq:expansion} are
\newcommand{\Y}{\mathcal{Y}}
\begin{align}
    \Y_{ab}^{\alpha\beta} \, \Y_{cd}^{\gamma\delta} \, \overline{\Y}^{bd}_{\beta\delta}  \epsilon^{xac} \epsilon_{\chi\alpha\gamma} \, , \notag\\
    \Y_{ab}^{\alpha\beta} \Y_{cd}^{\gamma\delta} \Y_{ef}^{\epsilon\zeta} \Y_{gh}^{\eta\theta} \epsilon^{xac} \epsilon^{deg} \epsilon^{bfh} \epsilon_{\chi\alpha\gamma} \epsilon_{\delta\epsilon\eta} \epsilon_{\beta\zeta\theta} \, , \notag\\
    \Y_{ab}^{\alpha\beta} \overline{\Y}^{ad}_{\alpha\delta} \overline{\Y}^{bf}_{\beta\zeta} \overline{\Y}^{gx}_{\eta\chi} \epsilon_{dfg} \epsilon^{\delta \zeta\eta} \, ,
    \Y_{ab}^{\alpha\beta} \overline{\Y}^{ad}_{\alpha\delta} \overline{\Y}^{bf}_{\epsilon\chi} \overline{\Y}^{gx}_{\beta\theta} \epsilon_{dfg} \epsilon^{\delta \epsilon\theta} \, . \label{eq:explicitContractions}
\end{align}

\bibliography{refs.bib}

\end{document}